\documentclass[twocolumn,nofootinbib,preprintnumbers,amsmath,amssymb]{revtex4}

\usepackage{graphicx}
\usepackage{dcolumn}
\usepackage{bm}
\usepackage{amsmath}

\def \be {\begin{equation}}
\def \ee {\end{equation}}

\begin{document}

\title{The collisionless hydrodynamics: On the nonexistence of collisionless shocks}

\author{Andrei Gruzinov}

\affiliation{ CCPP, Physics Department, New York University, 726 Broadway, New York, NY 10003
}

\date{January 28, 2025}

\begin{abstract}

  Collisionless shocks, essential for astrophysics, perhaps do not exist as statistically stationary solutions.  If so, any quantitative statement about a collisionless shock should be qualified by the age of the shock.

  A theoretical description of the upstream of the 1+1 dimensional electrostatic collisionless shock is developed -- {\it collisionless hydrodynamics}. Peculiarities of collisionless hydrodynamics prevent a shock formation when a piston is driven into cold plasma. An exact self-similar solution is found instead; the spatial extent of the solution grows linearly in time.

  Direct numerical simulations of plasma kinetics in 1+1 dimensions confirm the hydrodynamic result -- a statistically steady collisionless shock doesn't exist. Instead, at each fixed time, there is a continuous succession in space of marginally stable velocity distribution functions. The spatial support of this continuous succession grows linearly in time.

\end{abstract}

\maketitle

\section{Introduction: Collisionless Shocks in Astrophysics}

The mean free path of a $\sim 1{\rm GeV}$ proton or electron in $\sim 1{\rm cm}^{-3}$ plasma is $\gg 10{\rm Gpc}$, still, mildly relativistic blast waves propagate in the interstellar medium and cause gamma ray burst afterglows. Astrophysicists describe such blast waves using hydrodynamics; the front of the blast wave is described as an ideal zero-width hydrodynamic shock.

This approach can be justified if collisionless shocks exist, in the sense defined in \S\ref{S:E}. In \S\ref{S:H} we develop collisionless hydrodynamics, which appears to give a good description of the upstream of the simplest possible collisionless shock (1+1 dimensional, non-relativistic, electrostatic, electron-positron); we show that in this hydrodynamics the shock does not exist. We confirm the nonexistence of a shock by a direct numerical simulation in \S\ref{S:N}.  

What we see in the numerical simulation can be described as follows. As time goes on, the shock transition widens linearly in time, the width of the transition is $\sim Vt$, where $V$ is the velocity of the piston. At a fixed late time $t$ and a fixed position $x$, the velocity distribution function $f(t,x,v)$ is marginally stable to Langmuir waves (electrostatic plasma waves). As we change $x$ at a fixed $t$, we get a continuous succession of marginally stable velocity distributions. As we change $t$, the spatial support of this succession of marginally stable velocity distributions grows.

This scenario -- continuous succession of marginally stable velocity distributions with ever-widening spatial support -- is perhaps not limited to 1+1 dimensions, maybe it works in the physically relevant 1+3 case. Then theorists, when making any assertions regarding collisionless shocks, must study the dependence on the age of the shock. Perhaps, collisionless shocks should only be studied in those setting in which they appear in nature, in a collisionless blast wave or a bow shock.

\section{Strong Collisionless Shock}\label{S:E}

Consider the following setup: at $t=0$ there is cold electron-positron plasma of density $n_0$ in a half-space $x>0$ and then a perfectly reflecting planar piston  starts moving into the plasma, at non-relativistic velocity $V$, $x_{\rm piston}=Vt$. We want to find the resulting plasma flow at large $t$. We will solve this problem numerically and give a meaningful theory in 1+1 dimensions.

We want to solve the problem in 1+3 dimensions, but we can't. We can only predict, from dimensional analysis, the general form of the solution. For concreteness, consider a single quantity characterizing the flow, the width of the transition region $w$ across which the plasma density jump is occurring. In a collisionless regime (the plasma number density $n_0$, the elementary charge $e$, the electron mass $m$ enter only as the plasma mass and charge density $n_0m$, $n_0e$),
\be\label{E:w}
w(t)\sim \lambda \left(\frac{Vt}{\lambda}\right) ^\alpha,
\ee
where $\lambda$ is of order the Debye length of the shocked plasma
\be\label{E:lambda}
\lambda \sim \left(\frac{mV^2}{n_0e^2}\right)^{1/2}.
\ee

We say that the collisionless shock does not exist if the index $\alpha$ is not zero; then, on physical grounds, $\alpha >0$ and the transition width increases indefinitely. In this paper we show, both theoretically and numerically, that in 1+1 dimension $\alpha=1$.

The value of $\alpha$ in 1+3 dimensions, for a mildly relativistic velocity of the piston, is an interesting open problem. Eqs.(\ref{E:w}, \ref{E:lambda}) still apply, with $V$ replaced by $c$, but one needs to include the shock-generated magnetic field and use relativistic dynamics.

\section {Collisionless Hydrodynamics}\label{S:H}

We first repeat standard arguments for the existence of collisionless shocks in \S\ref{S:H1}, mainly because we need to present the plasma stability formalism. The plasma stability formalism is then used in \S\ref{S:H2} to deduce equations of collisionless hydrodynamics. In \S\ref{S:H3} we solve the piston problem in collisionless hydrodynamics.

\subsection {Plasma Instabilities}\label{S:H1}

Something interesting {\it must} happen when the piston moves into cold plasma. If nothing happens -- that is the plasma particles are just reflected by the piston -- the distribution function of electrons and positrons is
\be
\begin{split}
  f(t,x,v)= & n_0\theta(x-Vt)\\
  & \times\bigl( \delta(v)+\theta(2Vt-x)\delta(v-2V) \bigr). 
\end{split}
\ee

For
\be
t\gg \omega_p^{-1},~~~\omega_p^2\equiv \frac{8\pi n_0e^2}{m},
\ee
we can treat the region $Vt<x<2Vt$ as infinite and study linear perturbations of the form $e^{-i\omega t+ikx}$. Then, in Penrose's notations \cite{Penrose},
\be\label{E:dl}
k^2=Z(\frac{\omega}{k}),
\ee
~
\be
Z(\zeta)=\int\limits_{-\infty}^{+\infty}du~\frac{F'(u)}{u-\zeta}=\int\limits_{-\infty}^{+\infty}du~\frac{F(u)}{(u-\zeta)^2},
\ee
~
\be
F(u)=\omega_p^2g(u),
\ee
\be\label{E:deldel}
g(u)=\delta(u)+\delta(u-2V).
\ee.

In the frame of the piston, $g(u)=\delta(u-V)+\delta(u+V)$,
\be
Z(\zeta)=\omega_p^2\left( \frac{1}{(V-\zeta)^2}+\frac{1}{(V+\zeta)^2}\right),
\ee
and the dispersion law is
\be
\omega^2=\omega_p^2+V^2k^2 \pm \omega_p\sqrt{\omega_p^2+4V^2k^2}.
\ee
There is an (``two-stream'') instability for
\be
k<\sqrt{2}\frac{\omega_p}{V}\sim \frac{1}{\lambda}.
\ee
The fastest growing mode is
\be\label{E:imom}
\omega = \frac{i\omega_p}{2}, ~~~k=\frac{\sqrt{3}}{2}\frac{\omega_p}{V}.
\ee

The standard argument for the existence of a collisionless shock: the instability develops to nonlinear levels and assumes the role of collisions. During the instability growth time, $\sim \omega_p^{-1}$, a typical particle moves by $\sim V\omega_p^{-1}\sim \lambda$, so one assumes that the shock width is $w\sim \lambda$.

But the only rigorous theoretical statement we can make is Eq.(\ref{E:w}). This is because the growth rate Eq.(\ref{E:imom}) only works for strongly unstable distribution functions like Eq.(\ref{E:deldel}) -- such distributions are not found in an old, $t\gg \omega_p^{-1}$, shock. We must develop a nonlinear theory of collisionless shock.

\subsection {Equations of Collisionless Hydrodynamics}\label{S:H2}

Consider noninteracting particles in 1+1. The distribution function $f=f(t,x,u)$ satisfies the collisionless Boltzmann equation
\be
(\partial_t+u\partial_x)f=0.
\ee
It follows that
\be\label{E:hier}
\dot{I_k}+I_{k+1}'=0,
\ee
where the dot and the prime are $\partial_t$ and $\partial_x$, and $I_k(t,x)$ are the moments of the velocity distribution
\be
I_k(t,x)=\int du~u^kf(t,x,u).
\ee

For truly noninteracting particles, the infinite hierarchy of conservation laws, Eq.(\ref{E:hier}), is of no use. For interacting particles, one must add the fields or the potential energy, and the hierarchy Eq.(\ref{E:hier}) becomes invalid. But in some cases, like the ideal gas, the interactions do not change the first three equations of the hierarchy (particle number, momentum, and energy conservation, with particle mass equal to 1); the role of the interactions is only to give a unique shape to the velocity distribution function at a fixed particle number, momentum, and energy. Then the hierarchy is closed by the expression
\be\label{E:drh}
I_3=I_3(I_0,I_1,I_2).
\ee
For example, Maxwell distribution (in fact, any distribution with a vanishing third moment of the random part of velocity) gives
\be\label{E:sh}
I_3=\frac{3I_1I_2}{I_0}-\frac{2I_1^3}{I_0^2},
\ee
and the first three equations of the hierarchy become the conservation-law representation of (1+1)-dimensional hydrodynamics with the adiabatic index $\gamma=3$.

\subsubsection {Standard Hydrodynamics}

First suppose, incorrectly but usefully as a preparation for collisionless hydrodynamics, that standard hydrodynamics gives a fair description of the shock (only on length scales $\gg \lambda$, of course). As argued in \S\ref{S:H1}, rough applicability of standard hydrodynamics seems plausible. In the cold upstream (with density equal to 1), we have $I_0=1$, $I_k=0$ for $k>0$, which is in agreement with Eq.(\ref{E:sh}). It is not unthinkable that Langmuir waves are excited only in a narrow transition region, $\delta x\sim \lambda$, where the waves heat up the plasma to a roughly Maxwell distribution.

Then we have, assuming shock velocity equal to 1, in the shock frame, in the upstream,
\be
I_0=1,~~I_1=-1,~~I_2=1,~~I_3=-1.
\ee
In the shock frame, $\partial_t=0$, and, as follows from Eq.(\ref{E:hier}), $I_1,I_2,I_3={\rm const}$ -- constant fluxes of particles, momentum, and energy. With $I_1=-1,~I_2=1,~I_3=-1$, Eq.(\ref{E:sh}) reads
\be
-1=-\frac{3}{I_0}+\frac{2}{I_0^2},
\ee
or
\be
(I_0-1)(I_0-2)=0,
\ee
giving the postshock density $I_0=2$ (as it should be for $\gamma=3$).

The above is nothing more than standard hydrodynamics in the $I_k$ notations. In reality (meaning in the (1+1)-dimensional numerical experiment, \S\ref{S:N}), standard hydrodynamics does not describe collisionless plasma.

\subsubsection {Collisionless Hydrodynamics}

\begin{figure}
\includegraphics[width=0.4\textwidth]{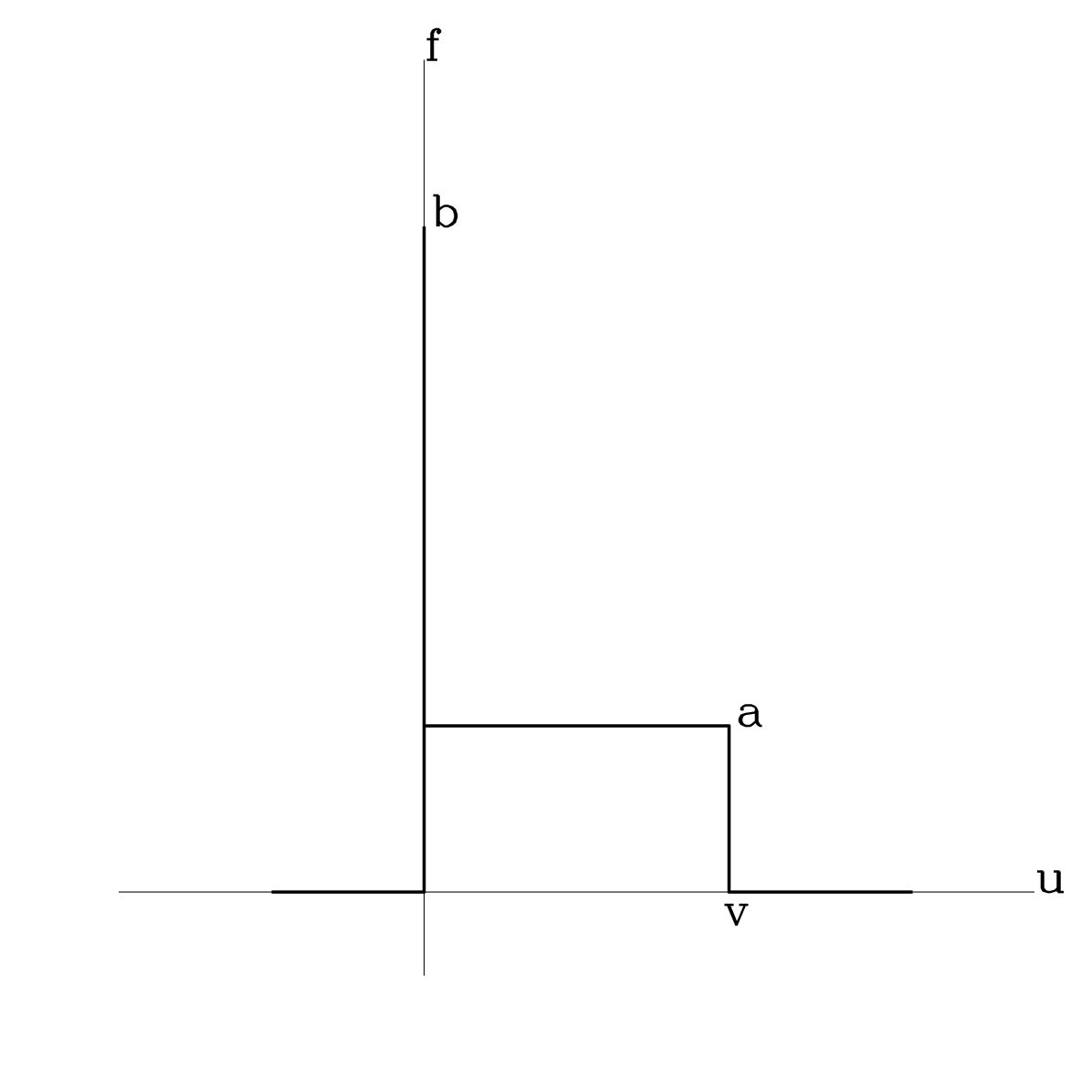}
\caption{\label{F:ft} Marginally stable velocity distribution $f(u)$: cold plasma density $b$, plateau $f=a$, extending to $u=v$.}
\end{figure}

Guided by the actual velocity distribution in the upstream, Fig.(\ref{F:fs}), assume the following 3-parametric velocity distribution, Fig.(\ref{F:ft}),
\be\label{E:ft}
f(t,x,u)=b(t,x)\delta(u)+a(t,x)\theta(u)\theta(v(t,x)-u).
\ee

Parameters $b$, $a$, $v$ are slow functions of $t$, $x$, meaning that they do not change much during $\delta t \sim \omega_p^{-1}$ or for $\delta x \sim \lambda$. The theoretical reason for the distribution Fig.(\ref{F:ft}) is its marginal stability and the availability of waves with phase velocities equal to nearly all velocities represented in the distribution $f$. Marginal stability means ${\rm Im}~\omega(k)=0$ for all $k$ and for all the three branches $\omega(k)$. The waves with the phase velocities all the way from $0$ to $v$ exist {\it iff} $av\ll b$, as is true in the far upstream. $\Big[$ {\it Proof:} The dispersion law, Eq.(\ref{E:dl}), reads
\be
Z(\zeta)=\frac{b}{\zeta ^2}+\frac{av}{\zeta (\zeta-v)}=k^2.
\ee
This gives a cubic equation for the phase velocity $\zeta\equiv\frac{\omega(k)}{k}$, so there are three branches $\omega(k)$. As $\zeta$ grows from $v+0$ to $+\infty$, $Z(\zeta)$ decreases from $+\infty$ to $+0$. It follows that there is exactly one real root $\zeta>v$ for any $k$. Analogously, there is exactly one real root $\zeta<0$ for any $k$. Since the cubic equation for the phase velocity has real coefficients, the third root is also real. The maximal phase velocity from the interval $0<\zeta<v$ is given by $Z(\zeta)=0$, which gives $\zeta=v/(1+av/b)$. $\Big]$

Langmuir waves with all phase velocities $\zeta$ from the interval $0<\zeta<v$ are required to exist and be marginally stable because in a weakly turbulent regime, which we assume,  only the resonant waves, $\zeta=u$, interact with the particles of velocity $u$ \cite{Vedenov}. Only the waves with $\zeta=u$ change the velocity distribution at $u$. Therefore the  waves must not be damped: ${\rm Im}~\omega(k)<0$ is not allowed. On the other hand, ${\rm Im}~\omega(k)>0$ is not allowed either, the instability would reshape the distribution function.

Any local change of the plateau distribution function, $f(u)={\rm const}$, would make the plasma unstable. This follows from the Penrose criterion \cite{Penrose} or, simpler and clearer, from the fact that the growth rate, if small, is proportional to the slope of the distribution function of resonant particles, ${\rm Im}~\omega(k)\propto \partial_uf(\frac{{\rm Re}~\omega(k)}{k})$ \cite{LL}.

Plateau distribution functions are well known in quasilinear theory \cite{Vedenov}. But quasilinear theory {\it does not} prove the ansatz Eq.(\ref{E:ft}). In quasilinear theory plateaus coexist with resonant Langmuir phonons \cite{Vedenov}, while we altogether ignore Langmuir waves. It is assumed, with no proof, that the Langmuir waves shape the distribution function without contributing to total energy, similar to hydrodynamics of ideal gases. As far as the author can see, Eq.(\ref{E:ft}) is no more than an assumption justified by numerical simulations; marginal stability and availability of resonant phonons is just an observation regarding the ansatz Eq.(\ref{E:ft}), not a proof of it. 

The defining relation of this new {\it collisionless hydrodynamics} is
\be\label{E:drch}
I_3=\frac{9}{8}\frac{I_2^2}{I_1},
\ee
as follows from 
\be\label{E:is}
I_0=b+av,~I_1=\frac{av^2}{2},~I_2=\frac{av^3}{3},~I_3=\frac{av^4}{4}.
\ee

Since Eq.(\ref{E:drch}) does not contain $I_0$, the momentum and energy equations form a closed system
\be\label{E:ch}
\begin{cases}
  \dot{I_1}+I_2'=0\\
  \dot{I_2}+\frac{9}{8}\left(\frac{I_2^2}{I_1}\right)'=0.
\end{cases}
\ee
Once $I_1$ is found, the density is given by the continuity equation
\be\label{E:ce}
\dot{I_0}+I_1'=0.
\ee

The first two equations of collisionles hydrodynamics, Eqs.(\ref{E:ch}), are similar (but not isomorphic) to ultrarelativistic (1+1)-dimensional hydrodynamics \cite{FGS}. Just like in \cite{FGS}, nonlinear Riemann invariants exist and give exact solutions of useful problems. 

\subsection {The Piston Problem in Collisionless Hydrodynamics}\label{S:H3}

We want to solve the piston problem in collisionless hydrodynamics: there is a perfectly reflecting piston at $x_{\rm piston}=Vt$; at time $t=0$, at all $x>0$, there is cold plasma of density $b=n_0$, and there is no hot plasma, $a=0$; find $b,a,v$ for $t>0$.

We show that the solution is not a shock (\S\ref{S:H3ns}), deduce boundary conditions at the piston (\S\ref{S:H3bc}), give an exact selfsimilar solution (\S\ref{S:H3ss}), compare the selfsimilar solution of collisionless hydrodynamics to direct numerical simulations (\S\ref{S:H3com}).

\subsubsection {There is no shock} \label{S:H3ns}

In collisionless hydrodynamics, a piston driven into cold plasma does not create a shock. $\Big[$ {\it Proof:} Suppose the shock travels with velocity $V_s$ relative to the cold unshocked plasma. The hydrodynamics  equations, Eqs.(\ref{E:hier}), then read
  \be\label{E:chs}
  -V_sI_k+I_{k+1}={\rm const},~~~k=0,1,2.
  \ee
In the unshocked plasma $I_1=I_2=I_3=0$, then Eqs.(\ref{E:chs}) give for the shocked plasma: 
  \be
  I_2=V_sI_1,~~~I_3=V_s^2I_1.
  \ee
But $I_3$ is also given by the defining relation of collisionless hydrodynamics, Eq.(\ref{E:drch}), so that
\be
V_s^2I_1=I_3=\frac{9}{8}\frac{I_2^2}{I_1}=\frac{9}{8}V_s^2I_1.
\ee
Then $I_1=0$ also in the downstream, and, from the first Eq.(\ref{E:chs}), $I_0={\rm const}$, meaning that there is no shock jump at all.  $\Big]$

\subsubsection {Boundary conditions} \label{S:H3bc}

To deduce boundary conditions at the piston, at $x=Vt$, consider the number of particles, momentum, and energy in front of the piston:
\be\label{E:id}
\begin{array}{l}
N=\int\limits_{Vt}^\infty dx~I_0, \\
P=\int\limits_{Vt}^\infty dx~I_1,\\
2E=\int\limits_{Vt}^\infty dx~I_2.
\end{array}
\ee
(Divergence of $N$ at the upper limit is irrelevant, we can always cut off the cold plasma density at large $x$ without any effect on the boundary conditions at the piston.)

We have
\be
\dot{N}=0,~~\dot{P}=F,~~\dot{E}=VF,
\ee
where $F$ is the force one needs to apply to move the piston at a constant velocity $V$. Now compute the rates of change $\dot{N}$, $\dot{P}$, $\dot{E}$ from the definitions Eqs.(\ref{E:id}) and the equations of motion Eq.(\ref{E:hier}):
\be\label{E:bci}
\begin{array}{l}
-VI_0+I_1=0, \\
-VI_1+I_2=F, \\
-VI_2+I_3=2VF.
\end{array}
\ee
Using Eqs.(\ref{E:is}), we get 
\be
\begin{cases}
  b=0, \\
  v(v-2V)=0.
\end{cases}
\ee
We must select $v=2V$, because $v=0$ gives $I_1=0$, and, by the first of Eqs.(\ref{E:bci}), $I_0=0$, which means that there is literally nothing at the piston. So, the boundary conditions at the piston are
\be\label{E:bc}
b=0,~~~ v=2V.
\ee

\subsubsection {Selsfsimilar solution} \label{S:H3ss}

Write collisionless hydrodynamics, Eqs.(\ref{E:ch}), in terms of (nonlinear) Riemann invariants
\be\label{E:ri}
\begin{cases}
  \dot{I_+}+vI_+'=0, \\
  \dot{I_-}+\frac{1}{2}vI_-'=0,~~~v\equiv\left(\frac{I_+}{I_-}\right)^{1/6},
\end{cases}
\ee
where
\be
I_-\equiv a,~~~I_+\equiv av^6.
\ee

Since $I_+$ propagates away from the piston, at large $t$ the solution is 
\be
I_+={\rm const}.
\ee
Then
\be
I_-=I_+v^{-6},
\ee
and the second Eq.(\ref{E:ri}) gives
\be
\dot{v}+\frac{1}{2}vv'=0,
\ee
with an exact solution
\be
v=\frac{2x}{t},
\ee
satisfying the right boundary condition
\be
v(t,Vt)=2V.
\ee

The correct value of the constant $I_+$ is decided by the yet-unused boundary condition at the piston, $b=0$. First we have to solve the continuity equation, Eq.(\ref{E:ce}), where $I_0$, $I_1$ are given by Eqs.(\ref{E:is}) with $a=I_+v^{-6}$ and $v=\frac{2x}{t}$. We get
\be
I_0=b+I_+\left(\frac{t}{2x}\right)^5,~~~I_1=\frac{1}{2}I_+\left(\frac{t}{2x}\right)^4,
\ee
then
\be
\dot{b}=-\frac{1}{32}I_+\frac{t^4}{x^5},
\ee
and then
\be
b=n_0-\frac{1}{160}I_+\left(\frac{t}{x}\right)^5.
\ee
For $b=0$ at the piston, at $x=Vt$, we must choose
\be
I_+=160n_0V^5.
\ee

The final result, in the piston frame, with the piston at $x=0$, reads
\be\label{E:sss}
\begin{array}{l}
b=n_0\left[1-\left(1+\frac{x}{Vt}\right)^{-5}\right], \\
a=\frac{5}{2}\frac{n_0}{V}\left(1+\frac{x}{Vt}\right)^{-6}, \\
v=2V\left(1+\frac{x}{Vt}\right).
\end{array}
\ee

The total plasma density is
\be\label{E:ssb}
n=b+av=n_0\left[1+4\left(1+\frac{x}{Vt}\right)^{-5}\right].
\ee

\begin{figure}
\includegraphics[width=0.4\textwidth]{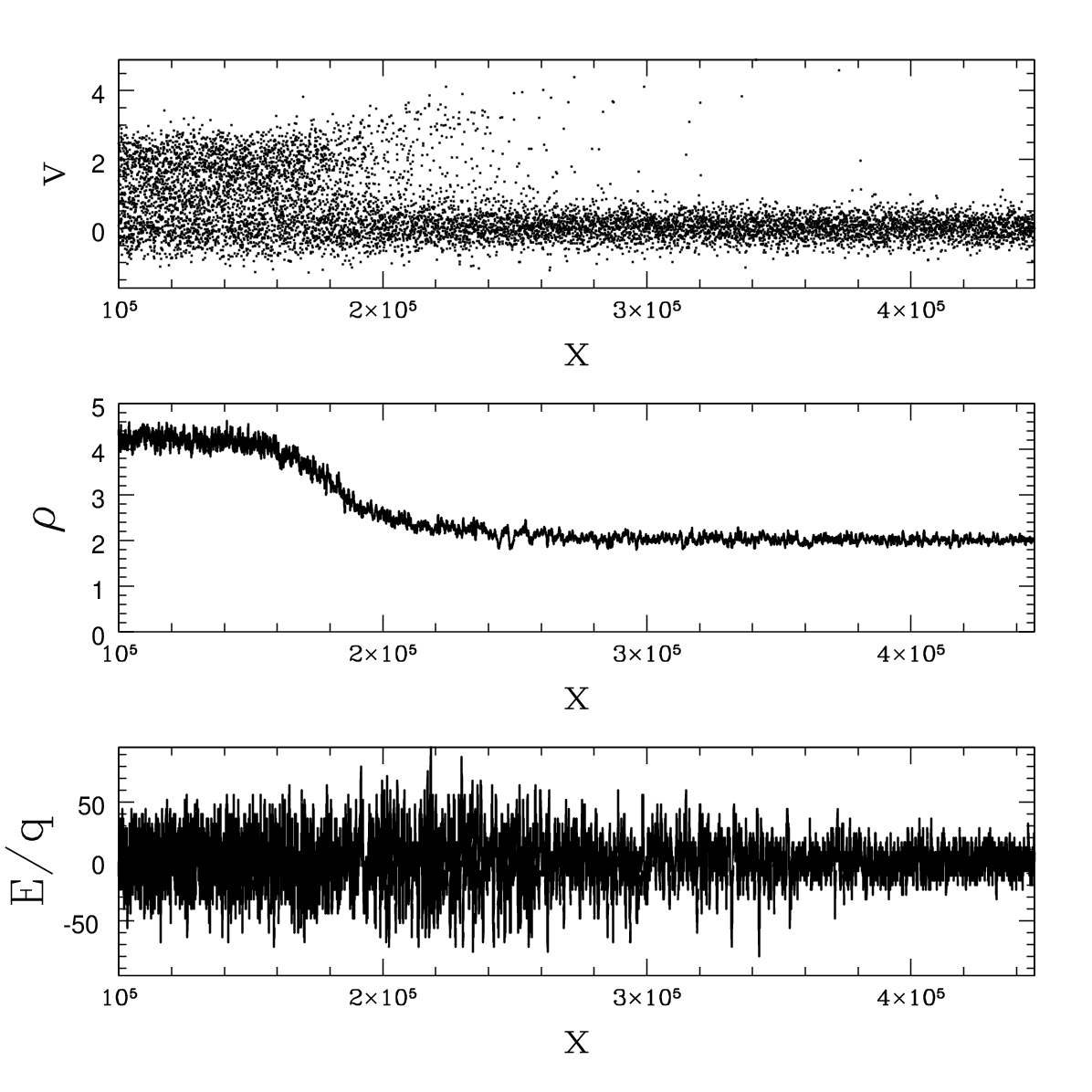}
\caption{\label{F:all} Particles in phase space, number density, electric field (divided by the elementary charge) at $t=10^5$.}
\end{figure}

\subsubsection {Comparison to numerics} \label{S:H3com}

Our solution fails near the piston. From Fig.(\ref{F:all}), the actual density at the piston is $n_{\rm piston}\approx 2.1n_0$, while our ``exact'' result is $5n_0$. Collisionless hydrodynamics fails near the piston because here the actual velocity distribution, Fig.(\ref{F:fp}), is very different from the ansatz, Fig.(\ref{F:ft}).

\begin{figure}
\includegraphics[width=0.4\textwidth]{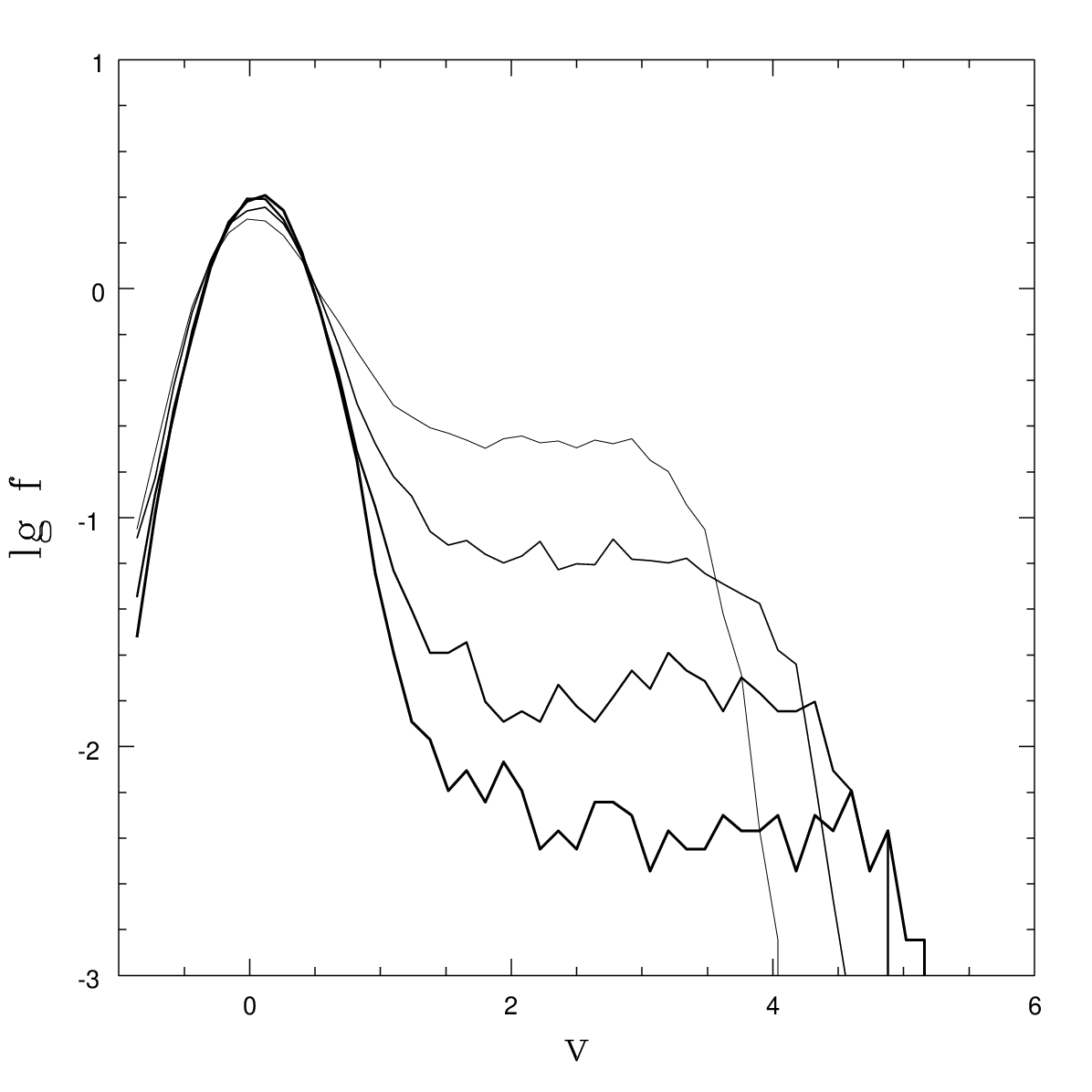}
\caption{\label{F:fs} Velocity distributions at $x=200000+35000\times k$, $k=0,1,2,3$, thicker lines correspond to larger $x$.}
\end{figure}

But in the upstream everything roughly works. The observed velocity distributions, Fig.(\ref{F:fs}), are close to the ansatz Fig.(\ref{F:ft}). The cutoff velocity $v$ does grow roughly linearly in space, as predicted by Eq.(\ref{E:sss}). The plateau value, $a$, does decrease roughly as predicted by Eq.(\ref{E:sss}).

 $\Big[$ But only ``very roughly''! For each $f$ shown in Fig.(\ref{F:fs}), draw the lowest horizontal line intersecting the graph of $f$ in two points with positive velocity coordinates. Measure the pair $(v,a)$ as the coordinates of the second intersection point. We get $v\approx 3.0,~3.5,~4.4,~5.0$ and we call this sequence roughly linear. The corresponding $a=0.2,~0.059,~0.013,~0.003$ have the ``best-fit'' power $a\propto v^{-8}$ and we call this roughly consistent with $a\propto v^{-6}$ of Eq.(\ref{E:sss}). $\Big]$

\begin{figure}
\includegraphics[width=0.4\textwidth]{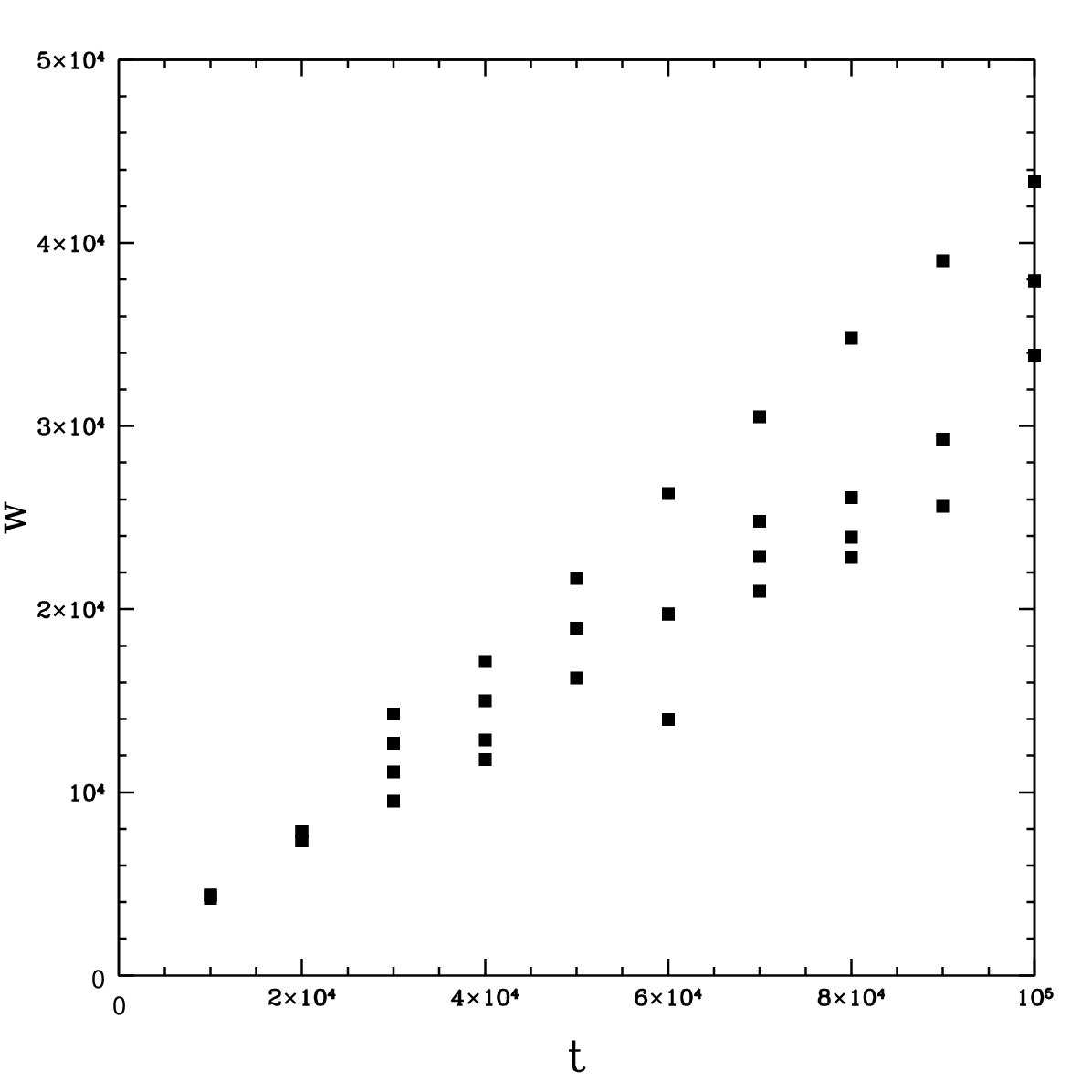}
\caption{\label{F:w} The width of the density transition region (where the number density $n$ grows from $(3n_{\rm min}+n_{\rm max})/4$ to $(n_{\rm min}+3n_{\rm max})/4$) as a function of time. Different values of $w$ at the same time $t$ are from different ways of computing the density $n(t,x)$.}
\end{figure}

Now estimate the theoretical width of the ``shock'' transition $w(t)$ as the width of the region where the total density jumps from $1.25n_0$ to $1.75n_0$ (as we do in \S\ref{S:N} when measuring the ``experimental'' width). From Eq.(\ref{E:ssb}) we then get
\be
w(t)\approx 0.34Vt,
\ee
in good agreement with Fig.(\ref{F:w}).

Although quantitative agreement with numerical simulations can't be claimed, collisionless hydrodynamics seems to make sense in the upstream.

\section{Conclusion}
 
    Collisionless hydrodynamics in 1+1 dimension, Eqs.(\ref{E:ch},\ref{E:ce}), is a ``well-posed'' theory.

    Collisionless hydrodynamics predicts that a piston pushed into cold plasma with velocity $V$ does not drive a shock, but rather a gradual transition of width $w(t)\approx 0.3Vt$, in rough agreement with a numerical simulation. In other words, collisionless shocks do not form in this 1+1 dimensional setup.

    Do collisionless shocks exist (as statistically steady solutions) in the real world? A good first step is to solve the mildly relativistic piston problem for a cold electron-positron upstream. Theoretically, the width of the transition is
\be
 w(t)\sim\delta \left(\frac{ct}{\delta}\right)^\alpha,
\ee
where $\delta\equiv\frac{c}{\omega_p}$ is the skin depth. The problem is to find the exponent $\alpha$. In our terms, $\alpha>0$ means that the shock does not exist.

\appendix

\section{Numerical Simulation}\label{S:N}

Electrostatic plasma in 1+1 dimension is easy to simulate numerically because the electric field is equal to the charge: for $x_{\rm piston}=Vt$, and charges $q_i$ at coordinates $x_i$, with $x_{\rm piston}<x_1<x_2<...$, the electric field acting on the particle $i$ is
\be
E_i=2\sum\limits_{j=1}^{i-1}q_j+q_i.
\ee
One needs to keep the particles ordered, but this is easy.  Only a small number of disorders appear at each time step; these can be removed by a simple bubble algorithm.

We use the piston velocity $V=1$, the unshocked plasma density $n_0=1$ (meaning $n_{\rm electron}=n_{\rm positron}=n_0=1$). We give a small temperature to the upstream, $<v^2>=0.09$ (just to enhance statistical averaging; we checked that smaller temperatures, $<v^2>=0.01$, give similar results). The particle mass is $m=1$, the charge is $q=0.0071$ so that the plasma frequency is $\omega_p=0.01$. 

In Fig.(\ref{F:all}) we show, at $t=10^5$, the particles in phase space, the plasma number density, and the electric field in the entire simulation box. If the electrostatic shock in 1+1 dimensions did exist, the width of the transition should have been $\sim \frac{V}{\omega_p} \sim 100$. The actual width of the transition is $\sim 30000$, and as shown in Fig.(\ref{F:w}) it grew linearly in time.

In Fig.(\ref{F:fs}) we show velocity distributions at different $x$ at $t=10^5$. These velocity distributions have been approximated by Eq.(\ref{E:ft}), Fig.(\ref{F:ft}). As one moves away from the piston: (i) the cold plasma density $b$ increases, (ii) the plateau value of the velocity distribution $a$ decreases, (iii) the cutoff velocity of the plateau $v$ increases. This is, qualitatively, what the collisionless hydrodynamics solution, Eqs.(\ref{E:sss}), predicts.

\begin{figure}
\includegraphics[width=0.4\textwidth]{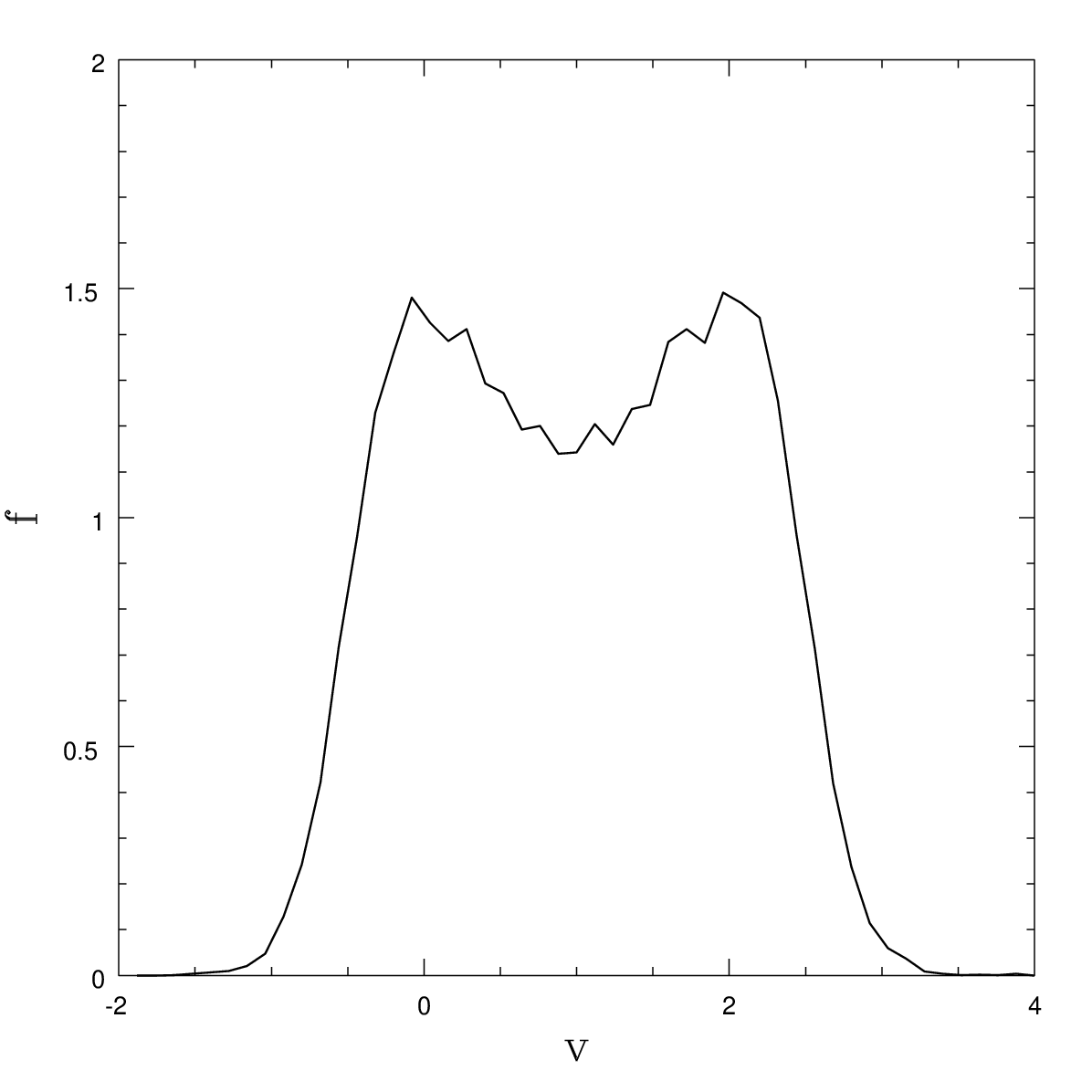}
\caption{\label{F:fp} Velocity distribution near the piston.}
\end{figure}

Finally, Fig.(\ref{F:fp}) shows the velocity distribution near the piston. Clearly this distribution cannot be approximated by Eq.(\ref{E:ft}), Fig.(\ref{F:ft}). Collisionless hydrodynamics description can only be valid in the upstream (if at all).

\end{document}